\documentclass[twocolumn,
aps,nofootinbib,showpacs,showkeys,preprintnumbers,
tightenlines] {revtex4}

\usepackage{ulem}

\newcommand{\dis}[1]{\begin{equation}\begin{split}#1\end{split}\end{equation}}

\newcommand{\etal}{{\it et al.}\ }

\newcommand{\GeV}{\,\mathrm{GeV}}

\newcommand{\keV}{\,\mathrm{keV}}

\def\mt2{M_{T2}}
\def\trialmt2{M_{T2}(\chi)}

\def\mthetat2{M_{\theta T2}}
\def\trialmthetat2{M_{\theta T2}(\chi)}

\def\mpit2{M_{\pi T2}}
\def\trialmpit2{M_{\pi T2}(\chi)}

\def\p0{{\bf p}_0}

\def\etal{{\it et al.}}

\usepackage{epsfig,subfigure,graphicx,amsmath,amssymb}
\usepackage{color}

\begin{document}
\title{Constraining WIMP magnetic moment from CDMS II experiment}
\author{Won Sang Cho$^{(a)}$, Ji-Haeng Huh$^{(a)}$, Ian-Woo Kim$^{(b)}$, Jihn E. Kim$^{(a)}$, and Bumseok Kyae$^{(a)}$}

\affiliation{$^{(a)}$Department of Physics and Astronomy and
Center for Theoretical Physics, Seoul National University, Seoul 151-747,
Korea}
\affiliation{$^{(b)}$Department of Physics, University of Wisconsin,
Madison, WI 53706, USA}
\begin{abstract}
We consider a degenerate or a nearly degenerate dark matter sector where a
sizable magnetic moment of an almost Dirac type neutral dark matter candidate $N$
is anticipated. Then, due to soft photon exchange,
the cross-section in direct detection of $N$ can be enhanced at low $Q^2$ region.
We discuss the implication of this type of models in view of the
recent CDMS II report.
\end{abstract}


\preprint{MADPH-10-1554}

\maketitle

\section{Introduction}
Recent cosmological and astrophysical observations strongly suggests
that a significant portion of energy density of the universe exists
in a form of dark matter (DM). DM has been speculated after the observation of the galaxy matter velocity distribution. The angular frequency distribution of the cosmic microwave background radiation, recent surveys of the
expansion rate of remote galaxies and the simulation of the structure formation support around 25\% of energy density in a form of non-relativistic matter while only one sixths of such matter has been identified as visible one.

Among many plausible theories of DM \cite{Axions} or equivalent modification
of gravity \cite{MOND}, weakly interacting massive particles (WIMP) are of particular interest \cite{WIMPs}. The scenario with WIMP assumes that DM particles were produced in the early Universe by thermal processes and the DM relic density has been frozen out by decoupling since that epoch. The DM energy density from the current observation is well matched with such a thermal production scenario with an electroweak scale DM mass and appropriate electroweak scale interactions. It is also very interesting if DM has some interesting connection with the origin of the electroweak symmetry
breaking as in the minimal supersymmetric standard model~(MSSM) \cite{ibanez}.
Above all, this WIMP scenario motivates searches for DM particles by
detecting and/or producing them directly in the laboratory.

Direct detection of DM has been carried out for several decades and the experimental sensitivity has been drastically improved in recent years \cite{Xenon10,CDMSII}. The abundant DM particles that flows through Earth may sporadically collide with ordinary matter nuclei, resulting in the recoil of a nucleus. The current WIMP search through the energy deposit in the cryogenic detector device has reached the DM cross-section with ordinary matter nuclei at the order of $\sim 10^{-44}$ cm$^2$, and  the upgraded CDMS II experiment \cite{CDMSII} has reached to this level.

In most DM models, fermionic DM is a Majorana particle, such as
neutralino of the MSSM \cite{Goldberg83}. However, we cannot rule out a Dirac fermionic DM $N$ at present, and hence it is very important to consider physical effects of such Dirac fermionic nature of DM. In this regard, we consider an appreciable magnetic 
moments of DM. In fact, the magnetic moments of neutral fermions were considered for a long time since the time of weak neutral currents \cite{numagmom} up to the present age of DM \cite{momDM}.

One notable feature in the scattering through magnetic moment of neutral fermion is that it has a  larger cross section for a lower momentum photon exchange.
The trend observed by two CDMS II candidates \cite{CDMSII} indeed show this behavior: the energy deposit at 12.3 keV and 15.5 keV (just above the threshold of 10 keV) while much larger energy deposit is allowed in that experiment.
At this time with two possible low $Q^2$ candidates of the CDMS II experiment \cite{CDMSII}, therefore, it is appropriate to scrutinize the Dirac DM aspect more closely. In particular, we will pay attention to the magnetic moment $f$ of $N$.\footnote{$f$ is the magnetic moment of $N$ in units of the $N$ Bohr magneton, $e/2m_N$.}
In principle, this study includes the effects of the electric dipole moment also, but we will not specify them explicitly which would have needed an additional assumption about CP violation.

There have been several works on the DM dipole moments~\cite{momDM,Masso09}. In these works, various observational constraints ({\it e.g.}
the cosmic microwave background radiation, the cosmic $\gamma$--ray detection,the DM relic abundance) have been considered. Although the authors of Ref.~\cite{momDM,Masso09} commented other possibilities, they mainly considered DM models in which the thermal DM relic density is determined only by annihilation due to magnetic dipole interactions. However, as pointed out by \cite{momDM,Masso09}, models that give rise to
the DM magnetic moment interaction with the photon exchange usually have other annihilation channels. Here, we consider the cases in which the relic density does not constrain the DM magnetic dipole moment. Nonetheless, DM direct detection
can constrain  the DM magnetic dipole moment.

As a prototype, we consider supersymmetic (SUSY) models with extra charged
singlets while direct interaction between DM and nuclei can be forbidden
in the leading order. The magnetic dipole moment is generically
induced by one-loop diagram. Such models have been discussed
in a certain class of leptophilic scenarios for explaining recent cosmic ray
anomalies, where DM particle has suppressed coupling with colored particles.

We also emphasize that our analysis fully includes nuclear anomalous magnetic
moment interactions. Without the photon exchange as in the neutralino case, the $F_2$ form factor effects of such nucleus are usually negligible since the range of the DM--nucleus interaction is much smaller than the size of nucleus. However,
the interaction through photon exchange can make the contribution comparable.
The nucleus magnetic moment effect has not been considered properly
in the previous works. Here, we present the DM detection rates including
such effects and compare them with the recent CDMS II data.

This paper is organized as follows.
In Sec. \ref{sec:MagMom},
we calculate the effective magnetic moment operator from a simple class of supersymmetric models. In generic models beyond the SM, one-loop induced magnetic moment should usually have a similar structure calculated in this section. In Sec. \ref{sec:decoupling}, we calculate the decoupling temperature of $N$ by solving the Boltzmann equation. We also use the $(g-2)_e$ bound to constrain the hypothetical  Yukawa coupling $\lambda$. In Sec. \ref{sec:Detection}, we calculate the direct detection cross-section and show the $Q^2$-dependence due to the magnetic moment, which can be observed by the recoil energy distribution. The behavior due to the magnetic moment is distinguishable from the predictions of other DM models. Next, the CDMS II experimental report is discussed in the context of DM models with a large magnetic moment. In this section, we also comment on the collider phenomenology of this class of models.  Sec. \ref{sec:Conclusion} is a brief conclusion, summarizing the allowed magnetic moment $f$ of Dirac or almost-Dirac DM models.

\section{Magnetic moment of $N$}\label{sec:MagMom}

\begin{figure}[!ht]
\begin{center}\resizebox{0.5\columnwidth}{!}
{\includegraphics{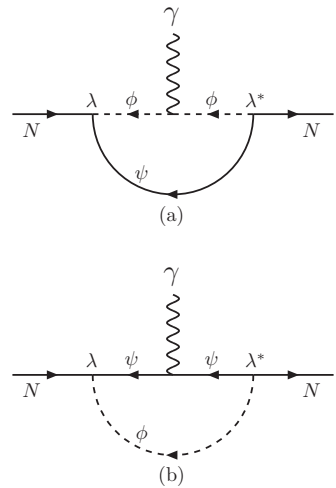}}
\end{center}
\caption{One-loop diagrams for the magnetic moment of $N$.}\label{fig:MagMom}
\end{figure}

A neutral Dirac fermion $N$ can acquire a magnetic moment as shown in Fig. \ref{fig:MagMom}
if it couples to charged particles, fermion $\psi$ and boson $\phi$, by a Yukawa coupling,
 $\lambda \bar\psi_R N_L\phi+$h.c.
In the MSSM, the lightest supersymmetric particle (LSP) is not a Dirac fermion and thus the LSP DM cannot acquire a magnetic moment.
However, in the extensions of MSSM with extra singlet chiral superfield, which have been proposed for many reasons : to address $\mu$-problem or to raise the mass of the lightest Higgs boson, the LSP $\chi$ of the MSSM sector and the additional singlet $N$ may form a single Dirac fermion state  \cite{HuhTwo,HuhDecaying}.

For a specific calculation, we consider the following
superpotential which has been discussed in \cite{HuhTwo}, \dis{
W=\lambda Ne^cE+m_N N\bar N +m_E EE^c+\rho N^3\, .
\label{eq:model} } In this model, there is no tree level
interaction between $N$ and nucleus because it couples only to
leptons and its scalar partners. For the magnetic moment, the
existence of the coupling of the type $\lambda Ne^cE$ is the
essential one, which can arise in many other models extending the
SM. For DM magnetic moment, it is required that $N$ must be a DM
component, presumably by an exact $Z_2$ symmetry. Actually Eq.(1)
is the simplest model in which  the magnetic moment can play an
essential role in the DM detection. More complicated cases can be
obtained by its proper extensions.

However, in the extension of MSSM,  the interaction type $\lambda \bar\psi_R N_L\phi+$h.c. is present through $NH_uH_d$ which however has to be avoided by the following reasons. Firstly, after the Higgs fields $H_u$ and $H_d$ develop VEVs, they give a VEV to $N$, which is harmful because of the mass mixing with the electron $e$. Second, it splits masses of $N$ and $\overline{N}$, and the anomalous magnetic moment interaction is a transition type between two mass eigenstates, and hence the direct detection rate as a function of recoil energy must be considered more carefully if the mass difference is of order 1--10 keV.

For a Dirac-type neutral fermion, its electromagnetic interaction
is dominated by the magnetic moment, as pointed out for the case
of neutrinos \cite{numagmom}. Let us define the magnetic moment of
$N$ as \dis{ \frac{ef}{2m_N}\overline{N}i\sigma^{\mu\nu} N
F_{\mu\nu}\, .\label{eq:defmmom} } For Eq. (\ref{eq:model}), $f$
is estimated from Fig. \ref{fig:MagMom}  \cite{Leveille},
\dis{
f= \frac{ |\lambda|^2m_N^2}{16\pi^2}\int_0^1  &dx\Big\{ \frac{q_\phi(x^2-x^3)}{m_N^2x^2+(m_\psi^2-m_N^2)x+m_\phi^2(1-x)}\\
&-
\frac{q_\psi(x^2-x^3)}{m_N^2x^2+(m_\phi^2-m_N^2)x+m_\psi^2(1-x)}\Big\}
\label{eq:gminustwo} }
where $m_\psi$ is the mass of the fermion ($e^c$ or $E$) and $m_\phi$ is the mass of the boson ($\tilde e^c$ or $\tilde E$) in the one-loop diagram. This calculation is an illustration for a sizable magnetic moment of a hypothetical DM particle $N$.
As shown here, a large magnetic moment is not unreasonable since $N$ is considered to be heavy.

If mass eigenstates splits the mass by a tiny amount, then the magnetic moment is of transition type with the initial $N$ and the final $N$ of Eq. (\ref{eq:defmmom}) considered different. Then, the lifetime of the heavier component is
\dis{
\frac{1}{\Gamma}=3.6\times 10^{-5}\left(\frac{10^{-6}}{f}\right)^2\left(\frac{m_N}{\rm 100~GeV}\right)^2\left(\frac{\rm 10~keV}{\Delta m_N}\right)^3 {\rm s}\label{eq:insidedecay}
}
where $\Delta m_N$ is the mass difference between two mass eigenstates of this almost-Dirac fermion. If the lifetime falls in the $10^{-10}$ s region with parameters chosen appropriately, then the decay products of $N$ deposit energy in the cryogenic detector. This happens for $f\le O(10^{-4})$ in which case the real signal from the cryogenic data must be revamped.

\section{Relic Density}\label{sec:decoupling}
Let us now proceed to discuss the effects of a large magnetic
moment of an extra singlet DM $N$ in cosmology and in particle
phenomenology. The DM relic density is given in terms of the
velocity averaged annihilation cross section ($\sigma_{\rm ann}
v$). The dominant annihilation channel of the Dirac DM $N$ allowed by Eq.~(\ref{eq:model}) is $N+\overline{N}\rightarrow e^-+e^+$, which is mediated by exchange of the scalar component
 of $E$.   It is straightforward to evaluate $\sigma_{\rm ann}v$,
which is approximately given by \dis{ \sigma_{\rm
ann}v=a+bv^2+{\cal O}(v^4). } $a$ and $b$ are \dis{
a=&\frac{3|\lambda|^4}{8\pi m_N^2(1+B)^2}\\
b=&\frac{|\lambda|^4}{48\pi m_N^2}\frac{(5B^2-16B-7)}{(B+1)^4},
}
where $B=m_{\tilde E}^2/m_N^2$.
Although there is also the annihilation by the magnetic dipole moment, Eq.~(\ref{eq:defmmom}), we will neglect it because its order of magnitude is estimated as
\begin{eqnarray*}
\sigma_{\rm ann}^{\rm dipole} v&\sim&\frac{1}{4\pi}\left(\frac{ef}{m_N}\right)^2\\
&\sim&\frac{|\lambda|^4}{4\pi m_N^2}\left(\frac{1}{16\pi^2}\frac{m_N^2}{{\rm Max}(m_\psi^2,m_\phi^2)}\right)^2\\
&<&10^{-4}\sigma_{\rm ann}v.
\end{eqnarray*}
where $\sigma_{\rm ann}$ is the $N\bar N$ annihilation cross
section.

\begin{figure}[!t]
\vskip 0.6cm
\begin{center}
\resizebox{0.8\columnwidth}{!}
{\includegraphics{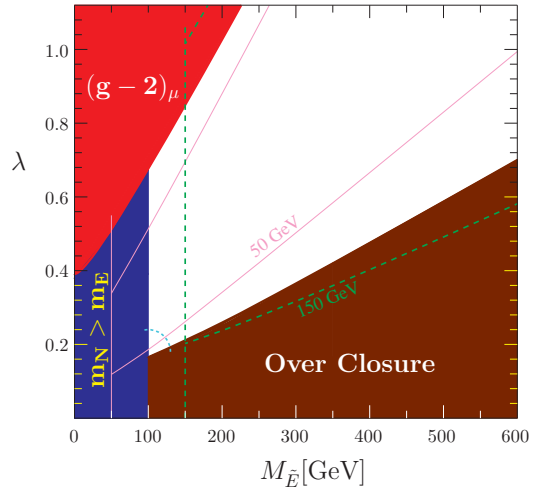}}
\end{center}
\caption{The allowed region in the $\lambda$ vs. $M_{\tilde E}$ plane from the relic density and $(g-2)_\mu$ bounds for $m_N=100$~GeV. The boundaries for $m_N=50$~GeV and 150~GeV are shown as the lavender line and the green dash line, respectively.}\label{fig:figRelic}
\end{figure}
In Fig. \ref{fig:figRelic} for $m_N=100$~GeV, we show the allowed
region from the constraint on the cosmic relic density and also
from the $(g-2)_\mu$ bound. The excluded regions are the lower
right wedge and the upper left wedge. Toward the $N$ DM scenario,
we also excluded the kinematically forbidden region $m_N>m_E$. For
the muon magnetic moment, we use the formula given in Eq.
(\ref{eq:model}) with $\mu^c N E$ and  Eq. (\ref{eq:gminustwo}).
The electron magnetic moment bound is buried in the muon magnetic
moment bound.  For the cases of $m_N=50$~GeV and 150~GeV, the
boundaries are shown as the lavender line and the green dash line,
respectively.

Since $f$ is of order $|\lambda|^2/16\pi^2\sim 0.6\times 10^{-2}|\lambda|^2$, the condition $|f|<10^{-4}$, as commented below Eq. (\ref{eq:insidedecay}), is satisfied only in a small skyblue dashed-arc region in the lower-left allowed region of Fig. \ref{fig:figRelic}. Except in this small area, we need not consider the possibility of a heavier component decay in the cryogenic detector.

\section{Direct Detection Rate}\label{sec:Detection}
\begin{figure}[t]
\begin{center}\resizebox{0.7\columnwidth}{!}
{\includegraphics{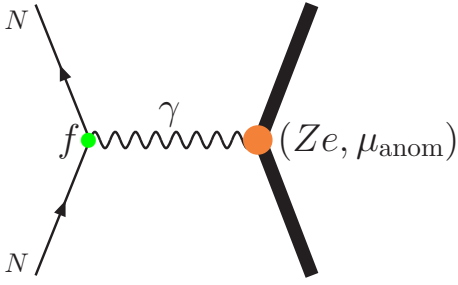}}
\end{center}
\caption{The elastic scattering of $N$ with a nucleus through the magnetic moment of $N$.}\label{fig:elastic}
\end{figure}

At low $Q^2$, the photon wavelength exceeds the nuclear size and
hence the whole nucleus, with charge and magnetic moment, acts as
a target. Thus, the elastic scattering cross section for the case
of a spin $\frac12$ nucleus target becomes, viz. Fig.
\ref{fig:elastic},
\begin{widetext}
\begin{eqnarray}
\frac{d\sigma}{dE_{\rm rec}}=\frac{2\pi\alpha_{\rm em}^2
f^2}{Mm_N^2|\vec p|^2}
&\bigg[&Z^2\left\{\frac{\Lambda_{-}(s,m_N^2,M^2)}{2ME_{\rm
rec}}+(2m_N^2+M^2-s) \right\} \nonumber\\
&+&2ZF_2(4m_N^2-ME_{\rm rec})+F_2^2\left\{
\frac{\Lambda_{+}(s,m_N^2,M^2)}{M^2}-\frac{2sE_{\rm rec}}{M}+E_{\rm
rec}^2\right\}\bigg],\label{eq:crosssec}
\end{eqnarray}
\end{widetext}
where $M$ is the mass of the nucleus with atomic weight $A$, $E_{\rm
rec}$ is the nuclear recoil energy, $Q_{\rm em}=Ze$ is the nuclear
charge,
\begin{eqnarray}
\Lambda_{\mp}(s,m_N^2,M^2)=(m_N^2+M^2-s)^2 \mp 4m_N^2M^2 ~,
\end{eqnarray}
and $F_2=\frac{1}{2}F^{\rm st}_2=\frac{1}{2}\left(\frac{\mu}{\mu_N}
\frac{m_{(Z,A)}}{m_p}-Z\right)$ for spin $\frac12$ case,
where $F^{\rm st}_2$ is the conventional notation. Our $F_2$ is a factor $\frac12$ of the conventional definition,  $F^{\rm st}_2$. The first two terms (`$Z$ terms') in the right hand side of Eq.~(\ref{eq:crosssec}) result
from the electromagnetic interaction between the tensor operator
of $N$ ($\overline{N}\sigma_{\mu\nu}N$) in Eq.~(\ref{eq:defmmom})
and the vector current of a spin $\frac12$ target nucleus
($\overline{\psi}\gamma_\mu\psi$).  On the other hand, the last
three terms (`$F_2$ terms') come from the interaction between the
tensor operators of $N$ and the nucleus. As seen in
Eq.~(\ref{eq:defmmom}), the tensor operator of a given nucleus
defines the spin of the nucleus
($\overline{\psi}\sigma_{\mu\nu}\psi$).

For the case of a higher spin target nucleus, `$\gamma_\mu$' and
`$\sigma_{\mu\nu}$' in the vector and tensor operators of the
nucleus should be replaced by larger dimensional representations
of the Dirac gamma matrix and the Lorentz generator, respectively
(see e.g. Ref.~\cite{weinberghighspin}).
Using Appendix, we can deduce that $F_2=\frac{1}{2}F^{\rm
st}_2=\frac{1}{2}\left(\frac{\mu}{\mu_N}
\frac{m_{(Z,A)}}{m_p}\sqrt{\frac{s_N+1}{3s_N}}-Z\right)$ from  Eq. (\ref{eq:crosssec}) with a higher spin generalization. In
the non-relativistic limit ($E_{\rm rec}\ll \{M,~m_N\}$ and $v\ll 1$), the differential cross section is given by
\begin{eqnarray}
\frac{d\sigma}{dE_{\rm rec}}&=&\frac{4\pi\alpha_{\rm em}^2
f^2}{m_N^2 E_{\rm rec}}\left[ Z^2(1-\frac{E_{\rm rec}}{2Mv^2}-\frac{E_{\rm rec}}{m_Nv^2})\right.\nonumber\\
&&\qquad\qquad\left.+
\left(\frac{\mu}{\mu_N}\right)^2\frac{s_N+1}{3s_N} \frac{ME_{\rm
rec}}{m_p^2v^2}\right].\label{eq:crosssec_nr}
\end{eqnarray}

\begin{figure}[!]
\vskip 0.6cm
 \begin{center}\resizebox{1.\columnwidth}{!}
{\includegraphics{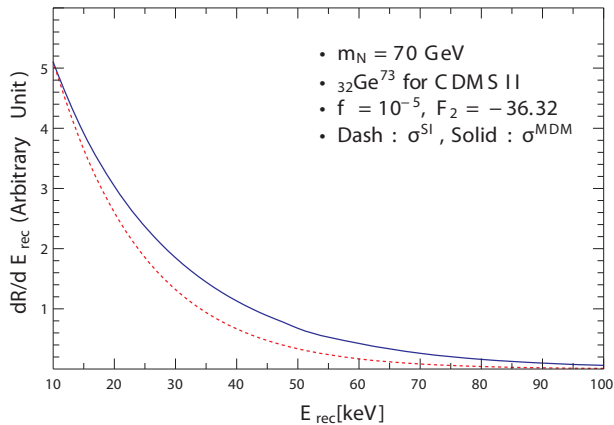}}
\end{center}
\caption{ WIMP($N$)--nucleus detection rate as a function of the recoil energy, normalized at $E_{\rm rec}=10$ keV. The solid line corresponds to the magnetic  moment and the dashed line corresponds to the SI interaction. The MDM graph and the SI graph
meet at two points due to our normalization to their equality at 10 keV. So, at very low $Q^2$, the MDM is bigger than SI as expected.
} \label{fig:MagMomScatt}
\end{figure}

The direct-detection rate (per unit detector mass) in a detector
with nucleus is given by
\begin{eqnarray}
\frac{dR}{dE_{\rm rec}}=\frac{\rho_N}{m_N M}\int_{|\vec{v}| >
v_{min}} d^3 \vec{v} f(\vec{v}) \frac{d\sigma}{dE_{\rm
rec}}.\label{eq:eventrate}
\end{eqnarray}
Here, we assume that the WIMP of mass $m_N$ accounts for the local DM density $\rho_N$ and have a local velocity distribution $f(\vec{v})$ with the normalization $\int d^3\vec{v}f(\vec{v})=1$. Using a simple Maxwell-Boltzmann velocity distribution
and $\rho_N\simeq 0.3 \GeV/{\rm cm}^3$, in Fig. \ref{fig:MagMomScatt}
we plot the expected direct-detection rates of an almost-Dirac DM $N$ (solid line) and the LSP $\chi$ (dash line) in the MSSM for the spin-independent(SI) interactions which always dominates for nuclei with $A\geq30$ in surveys of the SUSY parameter spaces \cite{Rosz}. The SI cross section is given by
\begin{eqnarray}
\frac{d\sigma}{dE_{\rm rec}}=\frac{2M}{\pi v^2}(Z f_p + (A-Z)f_n)^2
F^2(E_{\rm rec}),
\end{eqnarray}
where $f_p \simeq f_n \simeq 10^{-8}\GeV^{-2}$ are the SI couplings
of WIMPs to protons and neutrons, respectively, the SI form factor
$F(E_{\rm rec}) = 3 e^{-\kappa^2 s^2 /2}(\sin(\kappa r)-\kappa r
\cos(\kappa r))/(\kappa r)^3$, with $s=1$ fm, $r=\sqrt{R^2-5s^2}$,
$R=1.2A^{1/3}$ fm, and $\kappa=\sqrt{2m_n E_{\rm rec}}$.
\begin{table*}[!]
\begin{center}
\begin{tabular}{c|c|c|c|c|c|c|c|c}
\hline\hline &&&&\\[-1.1em]
& O(8,16)$^{3}$ & Na(11.23)$^{\frac{3}{2}}$ & Si(14,28)$^{2}$ & Ge(32,73)$^{\frac{9}{2}}$& I(53,127)$^{\frac{5}{2}}$ & Xe(54,131)$^{\frac{3}{2}}$ & Cs(55,133)$^{\frac{7}{2}}$
& W(74,183)$^{\frac{1}{2}}$
\\[0.2em]
\hline
$\mu/\mu_N$ & 1.66812 & 2.21752 & 1.1218 & $-0.879467$ & 2.81327 & 0.692 & 2.58 & 0.117785  \\[0.2em]
\hline
$F_2$& 4.83 & 13.36 & 4.02 & $-36.32$ & 94.56 & 6.52 & 83.93 & $-26.30$  \\[0.2em]
\hline
\end{tabular}
\caption{Magnetic moments $\mu$ of several target nuclei used in
cryogenic detectors.  Here, $\mu_N$ is the proton Bohr magneton
$\mu_N= e/2m_p$. The nuclear spin is denoted as superscripts. $F_2$ is the half of the conventional definition,  $F_2=\frac12 F_2^{\rm st}$, which is related to the effective anomalous magnetic moment corresponding to the equivalent particle that has the same mass, charge and magnetic dipole moment as the target nucleus. }
\label{tab:mgmm}
\end{center}
\end{table*}
In Fig. \ref{fig:MagMomScatt}, we present the differential event
detection rate for $m_N=70$ GeV, Eq. (\ref{eq:eventrate}), as a
solid line above the CDMS II threshold of about 10 keV. The
neutralino DM case with $m_\chi=70$ GeV is shown as a dashed line.
To show the recoil energy dependence, we choose an arbitrary
normalization such that both of these lines match at 10 keV. The
target is Ge with $Z=32$ and $A=73$.  Its anomalous magnetic moment
is $F_2=-36.32$ as shown in Table \ref{tab:mgmm}. The $1/E_{\rm
rec}$ dependence in the first term of Eq. (\ref{eq:crosssec})
generically leads the $dR^{MDM}/dE_{\rm rec}$ to diverge in low
$E_{\rm rec}$ limit. However, such a divergent effect is diminished
by small velocity of WIMP, $v^2\sim10^{-6}$ which appears in the
coefficient of the $1/E_{\rm rec}$-term as follows:
\dis{
\frac{\Lambda_{-}}{2M E_{\rm rec}}&=\frac{2 M |\vec{p}|^2}{E_{\rm
rec}}\simeq 2M^3  \frac{v^2}{E_{\rm rec}}\\
& \simeq2M^3\frac{10^{-6}}{10^{(-5\sim-4)}(\GeV)}.\label{eq:flatterQ2}
}
Due to the suppression of the IR divergent feature in
non-relativistic scattering, the $dR^{MDM}/dE_{\rm rec}$ divergently
larger than $dR^{SI}/dE_{\rm rec}$ only in the region below $E_{\rm
rec}\sim 10 \keV$, while the slope of the $dR^{MDM}/dE_{\rm rec}$
above $E_{\rm rec}\sim 10 \keV$ becomes eventually more flatter than
the SI. This is why the $dR^{MDM}/dE_{\rm rec}$ distribution looks
larger than $dR^{SI}/dE_{\rm rec}$ above $E_{\rm rec }~10 \keV$ in
Fig. (\ref{fig:MagMomScatt}). This non-relativistic $v^2$-
suppression in the $1/E_{\rm rec}$-term might result in further
interesting possibility. Due to the suppression, `$Z$ terms' can be
comparable to each other, even with `$F_2$ terms'. In this regard, we obtain the upper-bound on the DM magnetic moment
from the recent CDMS II data for both of the $F_2=0$ and $F_2\ne0$
cases. The ``maximal gap
method''\cite{maximalgapmethod} is used to estimate the proper
allowed region with a $90\%$ confidence level. The most stringent bound appears as $f\lesssim
2.88\times10^{-3}$ for $m_N\simeq 21$ GeV and $F_2\ne 0$. When we ignore the
contributions of nucleus anomalous magnetic moment, $F_2=0$, then
the upper-bound of $f$ is $1.16\times10^{-2}$ for
$m_N\sim100\GeV$. Taking into account the non-zero $F_2$, the
upper-bound goes down to $1.04\times10^{-2}$ for $m_N\sim 100\GeV$.
The event rate becomes larger with non-zero $F_2$ so that the
allowed region is more constrained, producing $1-10\%$ of difference
in $f$. However, we can easily expect that depending on the materials
in direct detection experiments, the difference can be significantly
amplified due to the enhanced magnetic dipole moment effect. Thus,
it is worthwhile to study the DM multi-pole interactions with nuclei
more carefully.
\begin{figure}[!]
\resizebox{0.9\columnwidth}{!}
{\includegraphics{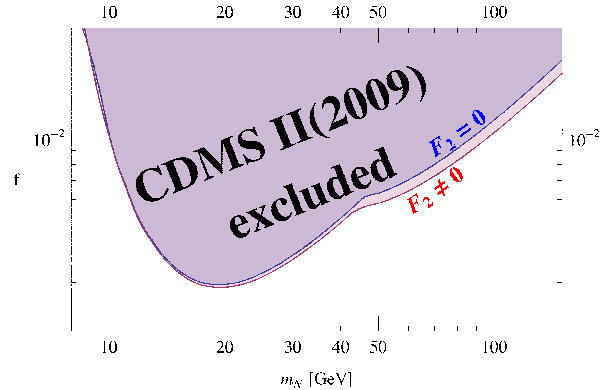}}
\caption{The allowed region of the DM magnetic moment $f$
vs. the DM mass $m_N$. The upper colored regions are excluded for $F_2=0$ and $F_2\ne 0$, respectively, with the 90 \% confidence level for the CDMS II data \cite{CDMSII}.}
\label{fig:MdmBoundRev}
\end{figure}

\section{Conclusion}\label{sec:Conclusion}
We considered a Dirac or an almost-Dirac DM $N$ which may acquire
a large magnetic moment. Using the possible dipole interactions, we
estimated the signal event rate which is expected in the
CDMS II experiment. Using the recent report of the CDMS II experiment, we present the upper-bound of the magnetic dipole moment.
The most stringent bound appears for $m_N\simeq 21$ GeV and $F_2\ne 0$:
 $f\lesssim 1.4\times10^{-3}$ with a sizable nucleus anomalous dipole
moment $F_2$ contribution. We point out that this sizable $F_2$ contribution is possible in the non-relativistic scattering of WIMP and nucleus, leading to a several factor improvement in the exclusion plot. In the future refined direct DM search experiments, the possibility of DM magnetic moment of $N$ can be probed with another independent information on its mass.

\vskip 0.5cm
\acknowledgments{
We thank J. H. Yoo for useful discussions.
WSC, JHH, JEK and BK are supported in part by the Korea Research Foundation, Grant No. KRF-2005-084-C00001. B.K. is also supported in addition by the FPRD of the BK21 program and the KICOS Grant No. K20732000011-07A0700-01110 of Ministry of Education and Science. IWK is supported by the U.S.
Department of Energy under grant No. DE-FG02-95ER40896.}
\vskip 0.5cm

\centerline{Appendix}
In the calculation of scattering cross-section of DM with target nuclei, we treated the nuclei as a spin $\frac12$ particle. However, many nuclei used in the direct detection of DM have spin different from $\frac12$. Moreover, our parametrization of the magnetic moment using $F_2$, in which spin  $\frac12$ for nuclei has been assumed, should be matched to produce a realistic value. For that, we consider two body wave function, in non-relativistic quantum mechanics with non-local interaction, which is obtained after integrating out the photon field,
\begin{equation}
\Psi^{\alpha i}(\vec x,\vec y)=N^\alpha(\vec x)\otimes\chi^i(\vec y)
\end{equation}
where $N^\alpha(\vec x)$ is the nuclear wave function with $S_z^{(N)}=\alpha$ and $\chi^i(\vec y)$ is the DM wave function  with $S_z^{(\chi)}=i$. The Scr\"odinger equation for this two body system reads
\dis{
H\Psi&=\left(-\frac{\nabla_x^2}{2M}-\frac{\nabla_y^2}{2m_N}\right)\Psi
+\frac{g_{\rm DM}Ze^2}{2m_N|\vec x-\vec y|}\nabla_x\cdot\vec S^{(N)}\Psi\\
&\quad +\frac{g_{\rm DM}g_{\rm nuclei}e^2}{2m_N2M|\vec x-\vec y|^3}\vec S^{(N)}\cdot \vec S^{(\chi)} \Psi\\
&=(H_0+H_{\rm mo\textrm -di}+H_{\rm di\textrm -di})\Psi\nonumber
}
where $H_{\rm mo\textrm -di}$ and $H_{\rm di\textrm -di}$ are the monopole--dipole and dipole--dipole interactions, respectively.
Here, we neglected unimportant normalization for each terms because we only
need to match the dependency on the nucleon spin.

Since the initial state of the problem is unpolarized, its density matrix
is proportional to the identity in $(2s_N+1)(2s_\chi+1)$-dimensional space,
that is, $\rho={\bf 1}/(2s_N+1)(2s_\chi+1)$, where $s_N$ and $s_\chi$ are the spin of nuclei and DM, respectively. Therefore, after averaging the
polarizations, the cross-section is proportional to ${\rm tr}[\rho M^\dagger
M]$, where $M$ is
\begin{equation}
M=H_{\rm int}+H_{\rm int}\frac{1}{E-H+i\epsilon}H_{\rm int}.
\end{equation}
For the dipole-dipole interaction contribution, the cross-section in the leading order becomes
\dis{
\sigma&\propto {\rm tr}[\rho M^\dagger M]\\
&\propto\frac{|g_{\rm DM}g_{\rm nuclei}|^2}{(2s_N+1)(2s_\chi+1)}
{\rm tr}\left[
\vec S^{(N)}\cdot\vec S^{(\chi)}\vec S^{(N)}\cdot\vec S^{(\chi)}
\right]\\
&=\frac{|g_{\rm DM}g_{\rm nuclei}|^2}{9}s_N(s_N+1)s_\chi(s_\chi+1)
} where the dot product is over the SO(3) space and Tr is over the
spin multiplicities. Here, if we use $g_{\rm
nuclei}=2M\mu/e\sqrt{s_N(s_N+1)}$, we can see that the term
proportional to $\mu^2$ is independent of the nucleus spin. Here, if we use $g_{\rm nuclei}={2 M
\mu}/{e s_N}$ we see that the term involving $\mu^2$ is
proportional to $(s_N+1)/3s_N$. It
confirms that our parametrization of $F_2$ in the text is
appropriate.

\newpage

\end{document}